# หลักสูตรคอมพิวเตอร์และความรู้ความสามารถทางอินเทอร์เน็ตของวิทยาลัยคอมพิวเตอร์ของเมืองอะกู

# Computer and Internet Literacy Course of the College of Computer Science for the Municipality of Agoo

*Clarisa V. Albarillo and others*

**บทคัดย่อ**

วัตถุประสงค์หลักของการศึกษาครั้งนี้ คือการให้ความตระหนักรู้ ความสามารถในการเรียนรู้และการพัฒนาทักษะด้านการใช้สารสนเทศต่อเจ้าหน้าที่ของ Agoo, La Union วัตถุประสงค์เฉพาะของการศึกษา คือ 1) เพื่อระบุลักษณะทั่วไปของผู้ตอบแบบสอบถาม เช่น ข้อมูลส่วนบุคคล ประวัติการศึกษา ความพร้อมและพื้นฐานในการใช้คอมพิวเตอร์ 2) เพื่อตรวจสอบประสิทธิภาพของหลักสูตรการรู้และการใช้คอมพิวเตอร์และอินเทอร์เน็ตด้านการให้บริการอย่างทันท่วงทีและการปรับปรุงความรู้ความสามารถทางคอมพิวเตอร์และอินเทอร์เน็ตของผู้เข้ารับการเรียนรู้ และ 3) เพื่อกำหนดระดับความเกี่ยวข้องของการฝึกอบรมของหลักสูตรการเรียนรู้และการใช้คอมพิวเตอร์และอินเทอร์เน็ต การศึกษาครั้งนี้เป็นการวิจัยเชิงบรรยาย เก็บข้อมูลโดยใช้แบบสอบถามเชิงสำรวจ วิเคราะห์ข้อมูลโดยใช้วิธีการทางสถิติ เช่น ความถี่ ร้อยละ และค่าเฉลี่ย ผลการศึกษาพบว่าผู้เข้ารับการเรียนรู้ส่วนใหญ่เป็นเพศหญิง คิดเป็นร้อยละ 88 โดยร้อยละ 84 ของผู้ตอบแบบสอบถามมีสถานะสมรส ร้อยละ 56 ของผู้ตอบแบบสอบถามมีอายุอยู่ในช่วงอายุ 30-39 ปี ในด้านของการศึกษา ร้อยละ 68 ของผู้ตอบแบบสอบถามจบการศึกษาระดับมัธยมศึกษาตอนปลาย และร้อยละ 84 ของผู้ตอบแบบสอบถามมีพื้นฐานทางด้านการใช้คอมพิวเตอร์ นอกจากนี้ผลการวิจัยแสดงให้เห็นว่าหลักสูตรการรู้และการใช้คอมพิวเตอร์และอินเทอร์เน็ตมีประสิทธิภาพในระดับมาก (4.67) ในด้านของการให้บริการ และมีระดับความความเกี่ยวข้องสูงกับเนื้อหาของหลักสูตร (4.45)

**คำสำคัญ:** หลักสูตรคอมพิวเตอร์และความรู้ความสามารถทางอินเทอร์เน็ต/ หลักสูตรความรู้ความสามารถ

ABSTRACT

The main objective of the study is to provide ICT awareness, literacy and skills development to the barangay officials of Agoo, La Union. Specifically, it aimed the following objectives: 1) to determine the profile of the respondents in terms of personal information, educational background and availability of computer unit and background in using computer; 2) to determine the effectiveness of the CILC in terms of services delivered, timeliness of the service, and improvement on the computer and internet knowledge of the trainees; and 3) to determine the level of relevance of the training sessions of the CILC. The study used a descriptive design. Data were gathered by using survey questionnaire and were analyzed by using statistical treatments such as frequency count, percentage and mean. As to the profile of the trainees, the study found that most of the trainees are female (88%); 84% are married, and 56% of them are at the age bracket of 30-39 years old. In terms of educational background, many are high school graduate (n= 17; 68%). In addition, most of them (84%) have background in computer.The result also shows that the CILC is at the high level of effectiveness (4.67) in terms of services delivered and is much relevant (4.45) in terms of its relevance.

Introduction

Behind the accomplishment of every educational institution lies in having achieved its vision, mission, goals, and objectives. The Don Mariano Marcos Memorial State University – South La Union Campus, College of Computer Science envisions itself in promoting computer literacy to the community it served, Over the years, the college is one of the leading institution in the Second District of La Union which positions itself as a leader in advocating the use of technology and its effective utilization in the one's daily life. The college started promoting computer literacy to the community specifically in the barangays of Agoo, La Union in the year 2002. A computer literacy training program for barangay officials, residents and out-of-school youth of the adapted barangay is conducted yearly by the college. It provides the participants an opportunity to be more productive and innovative in the service to their municipality and to the community. It also helps them develop their skills with the new technology in spite of not having finished any degree.

This is the overriding goal in the implementation of computer literacy training as an extension program of the college. This is anchored on the idea that to keep pace with the technological advancement and benefit from it, one must educate himself through computer literacy. Being computer literate gives the learner an opportunity to excel in rapidly changing environment in technology. According to Reynolds (2007), computer literacy means being knowledgeable about the capabilities of hardware and software and understanding how computers and the internet can enhance the learner's educational experiences.

Many academic and government institutions are now embracing the use of modern technologies in search for more efficient and effective process that can be used to improve both the quality of administrative functions and service to clients.

Governments and educational sectors across Asia Pacific are becoming increasingly concerned about the level of digital literacy. As today's society is becoming more and more dependent on new technology, increasing attention is given to computer literacy, which in the current information age is as significant as was reading, writing and calculus in the 19th-20th centuries (Anderson, 1983). Like reading, which is sometimes rightly called by socialization scholars - the socialization of socialization, or secondary socialization, computer literacy becomes an essential precondition for successful socialization and professional career. For this reason, education, being an important factor in society's development, plays an essential role in addressing the issue of digital literacy.

Computer literacy helps increase productivity in the work place. Everyone needs a computer to do basic tasks such as reports, documentation of records, etc. Computer literacy also helps to increase worker's value especially when they are trained at a higher level of computer skills. It also opens more doors to various opportunities to access greater resources.

In a study conducted, most of the people who are not familiar with using computers prior to the computer literacy training shared a lack of urgency towards learning the use of the same. However, after the course was taught to them, they showed keen interest in learning more and agreed it was beneficial to use the computer for several purposes.

In the international setting, among the 55 nations included in the Information Society Index (ISI), the Philippines is ranked 48th in terms of preparedness and ability to absorb advances and growth in information and communication technologies (ICT). Recognizing the need for improvements in the use of ICT in education and training, the Philippine Government has enacted laws to foster the use of ICT for widening access to education, improving the quality of teaching, and fostering the development of lifelong learning skills.

The Philippines could not be left behind in the application of ICT in everyday life. The Filipinos' penchant for anything that is new and western as well as the flare to adopt and be adept at new technology may have given us an edge in the use and mastery of ICT. The invention and popularity of "Chikka" as a social internet-communication tool, the tag of being the texting capital of the world and the proliferation of various ICT-based business process outsourcing schemes in the Philippines can only show how well the Filipinos could become experts in ICT if we choose to focus on it.

Due to the expanding embrace of ICT, its significance has gradually permeated into the areas of governance and public service. Though the trend may have started from among technologically advanced countries, the reach of ICT's relevance has undoubtedly spilled-over even into less developed countries. As a matter of fact, it has even become a crucial tool in order for developing countries to catch up with the trends of development in more progressive countries considering its borderless reach and application. The use of ICT is perceived to have elevated governance into a new level of effectiveness, efficiency and economy, which are the primordial goals of public administration. It has therefore assumed an indispensable role in every country's effort to keep phase with the global trend for development and make lives of people more convenient and productive. ICT, nonetheless, could not be a panacea to poverty and other ills of the society. It may just provide the means for a faster response to some immediate needs of the community.

Evidently, most of the local government in the Philippines had conducted series of trainings to educate the barangay folks specially the barangay officials in the field of Information and Communication Technology. As it was published in Mindanao Daily News (2013), the ICT Councils of Naga and Legaspi spearheads a series of free ICT Training workshops for the digitally uninitiated in Bicol. These ICT Councils join others under the National ICT Confederation of the Philippines in ensuring all Filipinos are digitally literate and can harness the good that technology can bring in improving lives. This national volunteer movement on digital literacy and citizenship is referred to as the DigiBayanihan movement.

On November 21, Out-of-School Youths and Barangay Champions in Naga, namely the Barangay Councilors, Barangay Tanods, Barangay Health Workers, Barangay Environmental Warriors, etc., will be trained on computer system familiarization and operation as well as typing speed and accuracy to be conducted by volunteer faculty and students from WRI Colleges.

While on November 22, selected First Year students of WRI Colleges may also learn about computer operations as well as more advanced IT skills such as photo editing, and PC troubleshooting.

To continually support this nationwide campaign of Computer Literacy and Proficiency Program, second and third batches will be happening on the months of January 2015 and March 2015, respectively.

On January 2015, to start the New Year and to empower people especially those who joined the first batch and on March 2015, to improve and commit those who volunteer, to spread the knowledge and learnings they gained on this movement.

In Legazpi, the Albay ICT Association, Inc. – Legazpi City ICT Council will spearhead a training on digital literacy for selected barangay officials on November 22. The training will provide much needed and relevant skills in digital literacy in order for barangay officials to be more efficient in providing public service.

Aimed at raising the level of digital literacy among the Filipino population, the workshops form part of DigiBayanihan, the unified efforts of the government, academe and IT industry to equip a greater numbers of Filipinos with 21st century skills that will enhance global competitiveness, workforce employability and entrepreneurship.

Likewise, with the initiative of the technical staff of the DILG Iloilo City, a Basic Computer (MS) Training for barangay secretaries in Arevalo District was conducted on June 4, 2015 held at the Conference Room A, 5th Floor of the New City Hall Building. The said training was seen important for the barangay secretaries in reinforcing their skills in handling basic Microsoft programs such as MS Work, MS Excel and MS Power Point. The barangay secretaries having the primary role in taking minutes and preparing reports and presentations in the barangay during sessions and in all the activities, requires competency in basic computer operation, thus the provision of the training.

Given the same objective to pursue a bold commitment towards the use of technology, the Department of Information and Communications Technology tapped the College of Computer Science to launch the Computer and Internet Literacy Course catered to the barangay officials of Agoo, La Union. The Department of Information and Communications Technology is an office existing by the virtue of the Executive Order No. 047 dated June 23, 2011 under the Department of Science and Technology. It is mandated to ensure the provision of efficient and effective information and Clarisa V. Albarillo and otherscommunications technology infrastructure to support client, effective transparent and accountable governance, and the speedy and efficient enforcement of rules and the delivery of accessible public services to the people.

As stated in the Local Government Code of the Philippines, Book III, Chapter 1, Section 384, as the basic political unit, the barangay serves as the primary planning and implementing unit of government policies, plans, programs, projects and activities in the community, and as a forum wherein a collective views of the people may be expressed, crystalized and considered, and where disputes may be amicably settled.

Barangay officials are deemed persons in authority in their respective jurisdiction, hence, they are mandated to maintain public order and ensure the protection of life, liberty and property. Thus, it is just but proper to educate the barangay officials on the latest trends in technology in order for them to keep abreast in latest technologies which can help them in their duties and responsibilities as public servants to the community they serve hence the Computer and Internet Literacy Course was launched.

The Computer and Internet Literacy Course has five areas such as Introduction to the Course (Basic Computer Concepts and Mouse and Keyboard Skills), Word Processing using MS Word, Internet, Email and Facebook, Spreadsheets using MS Excel, and Print and Preview Presentation using MS PowerPoint.

**Research Objectives**

The main objective of the study is to provide ICT awareness, literacy and skills development to the barangay officials of Agoo, La Union.

The study aimed at the following objectives:

1. To determine the profile of the respondents in terms of:
    a. Personal Information
    b. Educational Background
    c. Availability of computer unit and background in using computer;

2. To determine the effectiveness of the Computer and Internet Literacy Course in terms of:
    a.   Services delivered
    b.   Timeliness of the service
    c.   Improvement on the computer and internet knowledge of the trainees;
3. To determine the level of relevance of the training sessions of the CILC.

**Research Methodology**

This study employed the descriptive method of research.

Fluid Surveys (2014) articulated that descriptive research gathers quantifiable information that can be used for statistical inference on a target audience through data analysis. As a consequence, this type of research takes the form of closed-ended questions, which limits its ability to provide unique insights. However, when used properly, it can help an organization better define and measure the significance of something about a group of respondents and the population they represent.

Hence, this research determined profile of the respondents in terms of personal information, educational background and availability of computer unit and background in using computer and likewise determine effectiveness of the Computer and Internet Literacy Course in terms of services delivered, timeliness of the service, improvement on the computer and internet knowledge of the trainees and level of relevance of the training sessions of the CILC thus descriptive method of research was used.

The study was conducted at the Municipality of Agoo, La Union. This includes barangay officials of the different barangays of the municipality such as Capas, Purok, San Antonio, Sta. Barbara, San Joaquin Sur, San Juan, San Julian East, Sta. Monica, and San Nicolas Central.

Table 1 shows a total of 25 respondents involved in the study. It can be gleaned in the table that Capas, Purok, San Joaquin Sur, San Juan, San Julian East and Sta. Monica had only one representative while San Antonio and San Nicolas had two. Sta. Barbara had the most number of respondents totaling to 15. Total enumeration was used to maximize the accuracy of results.

**Table 1** Distribution of respondents

| Respondents | N |
|---|---|
| Capas | 1 |
| Purok | 1 |
| San Antonio | 2 |
| Sta. Barbara | 15 |
| San Joaquin Sur | 1 |
| San Juan | 1 |
| San Julian East | 1 |
| Sta. Monica | 1 |
| San Nicolas Central | 2 |
| Total | 25 |

Data gathering was done through identifying survey questionnaire to elicit the respondent's profile in terms of personal information, educational background and availability of computer unit and background in using computer and likewise the effectiveness of the Computer and Internet Literacy Course in terms of services delivered, timeliness of the service, improvement on the computer and internet knowledge of the trainees and level of relevance of the training sessions of the CILC.

The data gathered from the respondents were analyzed and interpreted by using statistical treatments as follows:

The researchers used frequency count and percentage in the profile of the respondents. Frequency counts was used to tally and describe the distribution of respondents according to personal information, educational background, and the availability of computer unit. Percentage was obtained by dividing the number of responses in every item by the total number of cases.

The quotient in its decimal form was multiplied by 100.

In order to determine the effectiveness of the Computer and Internet Literacy Course in terms of services delivered, timeliness of the service, the result of the administered survey questionnaire was construed through frequency count, computing for the mean and obtaining of the grand mean.

The obtained mean was interpreted through the use of a 5-Point Likert scale range of values and corresponding descriptive interpretations.

| Scale | Statistical Range | Descriptive Equivalent Rating(DER) |
|---|---|---|
| 5 | 4.20-5.00 | Very Much Effective |
| 4 | 3.41-4.19 | Much Effective |
| 3 | 2.60-3.39 | Moderately Effective |
| 2 | 1.80-2.59 | Slightly Effective |
| 1 | 1.00-1.79 | Not Effective |

To determine the improvement on the computer and internet knowledge of the trainees, the researchers recorded and compared the result of the respondents' scores during the pretest and posttest. The respondents who got a higher score during the posttest were said to have an improvement on their knowledge while those who had the same or lower scores during the posttest has not gained knowledge in the said projects.

To determine the level of relevance of the training sessions of the CILC, the result of the administered survey questionnaire was construed through frequency count, computing for the mean and obtaining of the grand mean. The obtained mean was interpreted through the use of a 5-Point Likert scale range of values and corresponding descriptive interpretations.

| Scale | Statistical Range | Descriptive Equivalent Rating(DER) |
|---|---|---|
| 5 | 4.20-5.00 | Very Much Relevant |
| 4 | 3.41-4.19 | Much Relevant |
| 3 | 2.60-3.39 | Moderately Relevant |
| 2 | 1.80-2.59 | Slightly Relevant |
| 1 | 1.00-1.79 | Not Relevant |

**Research Findings**

**Profile of the Trainees**

Table 2 shows the respondents' profile in terms of gender, civil status, and age.

In terms of sex, the female group dominated the trainees totaling to 22 or 88 percent of the whole population, while the male respondents totaled only 3 or 12 percent.

As to civil status, there are 84 percent who are married, 12 percent are single, and only four percent are widow(er). The statistics indicate that most of the respondents are married.

**Table 2** Distribution of Respondents according to Personal Profile

| Personal Information | Frequency (f) | Percentage (%) |
|---|---|---|
| Sex | | |
| Male | 3 | 12% |
| Female | 22 | 88% |
| Total | 25 | 100.00 |
| Civil Status | | |
| Single | 3 | 12% |
| Married | 21 | 84% |
| Widow(er) | 1 | 4% |
| Total | 25 | 100.00 |
| Age (years) | | |
| Below 21 y/o | 2 | 8% |
| 21 – 29 | 3 | 12% |
| 30 – 39 | 14 | 56% |
| 40 – 49 | 6 | 24% |

| Personal Information | Frequency (f) | Percentage (%) |
|---|:---:|:---:|
| 50 – 59 | 0 | |
| 60 y/o and | 0 | |
| Total | 25 | 100.00 |

Moreover, the table reflects that a great percentage of the respondents are at the age bracket of 30-39 years old with a frequency of 14 or 56 percent. The least number is in the age bracket of below 21 years old with a frequency of 2 or eight percent. This shows that the majority of the trainees are in their middle age which implies that most of them are enthusiastic in enhancing their computer skills and most of the barangay officials in this age bracket.

**Distribution of Respondents according to their Educational Background**

Table 3 exhibits the trainees' educational background. Many of the trainees are high school graduate (n= 17; 68%) while only one among the trainees is elementary graduate. This implies that most of the trainees did not enroll in any Bachelor's degree.

**Table 3** Distribution of Respondents according to their Educational Background

| Educational Attainment | Frequency (f) | Percentage (%) |
|---|:---:|:---:|
| Elementary Graduate | 1 | 4% |
| High School Graduate | 17 | 68% |
| College Undergraduate | 2 | 8% |
| College Graduate | 3 | 12% |
| Vocational Course | 2 | 8% |
| Total | 25 | 100.00 |

**Distribution of Respondents in terms of Availability of Computer Unit and Background in Computer**

Table 4 presents data on the availability of computer units at home, and the trainees' background in using computer.

It can be noted that majority (n =23, 92%) of the trainees has no available computer unit at home. Meanwhile, only 2 or eight percent of the trainees have computer unit at home.

Further analysis of the data in Table 3 indicates that 84 percent have background in computer and only 16 percent has no background in using it. This means that majority of the respondents already have a background on computers.

**Table 4** Distribution of Respondents in terms of Availability of Computer Unit and Background in Using Computer

| Availability of Computer Unit | Frequency (f) | Percentage (%) |
|---|---|---|
| Computer is available at home | | |
| With | 2 | 8% |
| Without | 23 | 92% |
| Total | 25 | 100.00 |
| Background in Using Computer | | |
| With | 21 | 84% |
| Without | 4 | 16% |
| Total | 25 | 100.00 |

**Effectiveness of the Computer and Internet Literacy Course**
**Services Delivered and Timeliness of the Service**

Table 5 discloses the result on the level of effectiveness of the CILC in terms of services delivered and the timeliness of the service. Both of the category yield a very much effective descriptive equivalent rating. The result implies that the service delivered by the trainors and facilitators is very much effective and is suited to the learning needs of the trainees thus contributed to the improvement of their knowledge and skills in using computer, as implied in the mean of 4.76.

Table 5 Level of Effectiveness of the Computer and Internet Literacy Course in terms of Services delivered and Timeliness of the service

| Effectiveness of the Computer and Internet Literacy Course in terms of: | Mean | Descriptive Equivalent Rating |
|---|---|---|
| Service Delivered | 4.76 | Very Much Effective |
| Timeliness of the Service | 4.57 | Very Much Effective |
| AWM | 4.67 | Very Much Effective |

**Knowledge of the Trainees**

Table 6 discloses the improvement on the knowledge of the trainees in the five projects which are Introduction to the Course, Word Processing using MS Word, Internet, Email and Facebook, Spreadsheets using MS Excel and Print and Preview using MS PowerPoint.

It can be gleaned on the table that in the Introduction to the Course, majority of the trainees (n=22; 88%) got higher score during the posttest which means that they have learned knowledge in the said project.

Table 6 Improvement on the Knowledge of the Trainees during the Post test

| Level of Improvement | Frequency (f) | Percentage (%) |
|---|---|---|
| Introduction to the Course | | |
| Got higher score | 22 | 88% |
| Retain | 1 | 4% |
| Got lower score | 2 | 8% |

Clarisa V. Albarillo and others| Level of Improvement | Frequency (f) | Percentage (%) |
|---|---|---|
| Total | 25 | 100% |
| **Word Processing using MS Word** | | |
| Got higher score | 25 | 100% |
| Total | | |
| **Internet, Email and Facebook** | | |
| Got higher score | 23 | 92% |
| Retain | 2 | 8% |
| Total | 25 | 100% |
| **Spreadsheets using MS Excel** | | |
| Got higher score | 25 | 100% |
| Total | | |
| **Print and Preview using MS PowerPoint** | | |
| Got higher score | 25 | 100% |
| Total | 25 | 100% |

As to the Internet, Email and Facebook project, most of the respondents with a frequency of 23 and a percentage of 92 got higher score during the posttest. This implies that the level of knowledge of the trainees were augmented after the hands on training.

In a lighter note, all of the trainees got higher score during the post test in the projects Word Processing using MS Word, Internet, Spreadsheets using MS Excel and Print and Preview using MS PowerPoint. This implies that the knowledge of the trainees improved.

**Level of Relevance of the Computer and Internet Literacy Course**

Table 7 presents the assessment of the trainees to the level of relevance of the training sessions of the CILC namely Introduction to the Course, Word Processing using MS Word, Internet, Email and Facebook, Spreadsheets using MS Excel, and Print and Preview using MS PowerPoint.

The trainees assessed the training session in Word Processing using MS Word to be very much relevant and had the highest mean rating of 4.76. It can be deduced that the trainees find the training session in Word Processing using MS Word to be very much relevant hence they can use the knowledge and skills that they have acquired in the said literacy course to their respective functions as barangay officials most especially in preparing reports, resolutions and other pertinent papers related to their functions.

Meanwhile, the training session on Internet, Email and Facebook got the lowest mean rating of 4.14 but still considered as much relevant. Such result may be attributed to the unfamiliarity of the trainees with the use of Internet, and only few of them have email addresses and facebook accounts due to unavailability of internet connection and devices.

In totality, the trainees evaluated the five training sessions as much relevant hence the said literacy training has served its purpose to heighten the knowledge and skills of the trainees in using that it suits to the needs of the trainees.

Table 7 Level of Relevance of the Training Sessions of the Computer and Internet Literacy Course

| Relevance of the Training Sessions of CILC | Mean | Descriptive Equivalent Rating |
|---|---|---|
| Introduction to the Course | 4.24 | Much Relevant |
| Word Processing using MS Word | 4.76 | Very Much Relevant |
| Internet, Email and Facebook | 4.14 | Much Relevant |
| Spreadsheets using MS Excel | 4.57 | Very Much Relevant |
| Print and Preview using MS PowerPoint | 4.52 | Very Much Relevant |
| AWM | 4.45 | Much Relevant |

**Conclusions and Discussions**

Based on the findings of this study, the following conclusions are forwarded:

1. Most of the respondents are females, married and come from the age bracket of 30-39 years old. In addition, majority of respondents are high school graduates, and have background in using computer.

2. The said CILC is very much effective thus most of them improved their knowledge on the different applications taught during the training course.

3. The said training course is much relevant as it suits and applies to the needs of the trainees.

**References**

Anderson, C. H. (1983). Computer literacy: Rationale, definition and practices. *Paper presented at a Satellite Teleconference on Microcomputers in Education.* ERIC ED228983.

Fluid Surveys (2014). *Descriptive research: Defining your respondents and drawing conclusions.* Retrieved from http://fluidsurveys.com/university/descriptive-research-defining-respondents-drawing-conclusions/.

Mindanao Daily News (2013, April 3). *Retrieved from http://issuu.com/sudaria_publications/docs/mdnapr3/5.*

Reynolds, C. R. (2007). *Encyclopedia of special education: A reference for the education of children, adolescents, and adults with sisabilities and other exceptional individuals,* New Jersey: John Wiley & Sons, Inc.

**Authors**

**Clarisa V. Albarillo,** Don Mariano Marcos Memorial State University
E-mail: clarisa.albarillo@dmmmsu-sluc.com

**Emely A. Munar,** Don Mariano Marcos Memorial State University
E-mail: emely.munar@dmmmsu-sluc.com

**Maria Concepcion G. Balcita,** Don Mariano Marcos Memorial State University
E-mail: mariaconcepcion.balcita@dmmmsu-sluc.com